\documentclass[pdftex,epjc3]{svjour3}                     
\smartqed  
\usepackage{graphicx,url}
\journalname{Eur. Phys. J. A}
\usepackage{physics} 

\usepackage{multirow}
\usepackage{booktabs}
\usepackage{placeins}

\usepackage{color}

\newcommand{\unit}[1]{\;\mathrm{#1}}
\newcommand{\iu}{\mathrm{i}}
\newcommand{\me}[1]{\mathrm{e}^{#1}}
\newcommand{\dvv}[3]{\frac{\mathrm{d}^2#1}{\mathrm{d}#2\,\mathrm{d}#3}}

\def\be{\begin{equation}}
\def\ee{\end{equation}}
\def\ba{\begin{align}}
\def\ea{\end{align}}

\def\lsim{\raise0.3ex\hbox{$\;<$\kern-0.75em\raise-1.1ex\hbox{$\sim\;$}}}
\def\gsim{\raise0.3ex\hbox{$\;>$\kern-0.75em\raise-1.1ex\hbox{$\sim\;$}}}

\def\i{{\rm i}}

\def\e{{\rm e}}

\renewcommand{\vec}[1]{\boldsymbol{#1}}

\begin{document}

\title{Alternative coalescence model for deuteron, tritium, helium-3 and their antinuclei}

\author{M.~Kachelrie{\ss}$^{1}$ \and S.~Ostapchenko$^{2,3}$ \and J.~Tjemsland$^{1}$}

\institute{$^1$ Institutt for fysikk, NTNU, Trondheim, Norway
  \and
 $^2$ Frankfurt Institute for Advanced Studies, Frankfurt, Germany
 \and
 $^3$ D.V.~Skobeltsyn Institute of Nuclear Physics,
 Moscow State University, Moscow, Russia}

\date{Received: date / Accepted: date}

\maketitle

\begin{abstract}
Antideuteron and antihelium nuclei have been proposed as a detection channel
for dark matter annihilations and decays in the Milky Way, due to
the low  astrophysical background expected. To estimate both the signal for
various dark matter models and the astrophysical background, one employs
usually the coalescence model in a Monte Carlo framework. This allows one
to treat the production of antinuclei on an event-by-event basis,
taking thereby into account momentum correlations between the antinucleons
involved in the process. This approach lacks however an
underlying microscopic picture, and the numerical value of the coalescence
parameter obtained from fits to different reactions varies considerably.
Here we propose instead to combine event-by-event Monte Carlo 
simulations with a microscopic coalescence picture based on the Wigner function
representations of the produced antinuclei states. This approach allows us
to include in a semi-classical picture both the size of the formation region,
which is process dependent, and momentum correlations. The model contains
a single, universal parameter which is fixed by fitting the production spectra
of antideuterons in proton-proton interactions, measured at the Large Hadron Collider. 
Using this value, the model describes well the production of various
antinuclei both in  electron-positron annihilation and in proton-proton
collisions.
\end{abstract}

\maketitle

\section{Introduction} \label{sec:intro}

Antideuteron and antihelium nuclei have been suggested as promising detection 
channels for dark matter, because of the low astrophysical background
expected for such signatures~\cite{Donato:1999gy}:
The dominant background source of  antideuterons are cosmic ray protons
interacting with the interstellar medium. The high threshold energy
for this reaction channel implies that the antideuterons produced by
cosmic rays have relatively large kinetic energies. Low-velocity antideuterons
are therefore an ideal tool to search for exotic sources of antimatter.
In the case of antihelium nuclei, the suppression of astrophysical backgrounds
at low velocities is even stronger, but the maximal event rates expected in
dark matter models are challenging for square-meter sized detectors.
At present, the search for antinuclei is performed by the AMS-02 experiment
on board of the International Space Station, while the GAPS balloon experiment
is planned to fly in the next Solar minimum period around 2020 or
2021~\cite{Battiston:2008zza,Aramaki:2015laa}.

The production of light clusters of antinuclei like antideuteron,
antihelium or antitritium\footnote{Since our discussion applies
  equally well to the production of particles and of antiparticles in $pp$ 
  and $e^+e^-$ collisions, the preposition `anti' is dropped further on.} is
usually described in the context of coalescence~\cite{Schwarzschild:1963zz,butler,Sun:2015jta,Zhu:2015voa,Zhu:2017zlb} or of statistical-thermal
  models~\cite{Acharya:2017bso,Andronic:2017pug,Vovchenko:2018fiy,Bellini:2018epz,Chen:2018tnh,Xu:2018jff,Oliinychenko:2018ugs}.
  In coalescence models,
 cluster formation has been traditionally parametrised by an invariant
coalescence factor $B_A$ which relates the invariant yield
$E_A\dv*[3]{N_A}{P_A}$ of nuclei with mass number $A$ formed out of $Z$
protons and $N$ neutrons to the invariant yields  $E_i\dv*[3]{N_i}{P_i}$
of protons ($i=p$) and neutrons  ($i=n$) via
\be
 E_A \dv[3]{N_A}{P_A} =
 B_A \left( E_p \dv[3]{N_p}{P_p} \right)^Z
 \left. \left( E_n\dv[3]{N_n}{P_n} \right)^N \right|_{P_p=P_n=P_A/A} .
\ee
In $e^+e^-$ and $pp$
collisions, one imposes typically the coalescence condition in momentum
space, requiring that the momenta of merging  nucleons in their
two-body center-of-mass (CoM) system are smaller than
some critical value $p_0$. In the limit of isotropic and equal proton and
neutron yields, the so-called coalescence momentum $p_0$ is related to
$B_A$ via
\be
B_A = A \left( \frac{4\pi}{3}\,\frac{p_0^3}{m_N}\right)^{A-1} ,
\ee
where $m_N$ denotes the nucleon mass. This scheme can be improved taking
into account the momentum correlations between nucleons, which are provided
by Monte Carlo (MC) simulations on an event-by-event basis.  Such an approach,
which was first suggested in Refs.~\cite{Kadastik:2009ts,Dal11}, is commonly
used for the prediction of
the antideuteron yield both from dark matter annihilations or decays and
from cosmic rays interactions~\cite{Cui:2010ud,Ibarra:2012cc,Fornengo:2013osa,Dal:2014nda,Grefe:2015jva,Herms:2016vop,Korsmeier:2017xzj,Coogan:2017pwt,Poulin:2018wzu},
for a review see Ref.~\cite{Aramaki:2015pii}. 
The only free parameter of this model is the coalescence momentum $p_0$,
which should be independent of the reaction type and the center-of-mass
energy $\sqrt{s}$
in order to be predictive. However, the numerical value of the coalescence
parameter obtained from fits to different reactions varies
considerably~\cite{Ibarra:2012cc,Aramaki:2015pii}.

An alternative scheme was developed to describe the formation of light
nuclear clusters in heavy-ion collisions. There, the
coalescence condition was imposed in coordinate space,  assuming that the
coalescence factor $B_A$ of a cluster with mass number $A$
is proportional to $V^{A-1}$, where $V$ denotes the volume of the
emission region of hadrons from the expanding cloud of
partons~\cite{Csernai:1986qf,Nagle:1996vp}.
There have been considerable efforts to combine these two approaches and
to develop coalescence models which are based on a microscopical
picture using, e.g., Wigner functions~\cite{Sato:1981ez}
  or a diagrammatic approach~\cite{Duperray:2002pj}.
Many of these attempts impose the coalescence condition in phase space,
using either a classical or quantum mechanical description as a starting point.
Such models have been mainly applied to heavy ion collisions and are reviewed,
e.g., in Refs.~\cite{Danielewicz:1991dh,Scheibl:1998tk,Chen:2018tnh}. An
interesting application of this approach to the prediction of antideuteron
and antihelium production by cosmic rays has been made recently in
Ref.~\cite{Blum:2017qnn}.

The coalescence process has also been modelled as a dynamical process where
the formation probability of a deuteron is proportional to the scattering
cross section of the reaction $\bar{N}_1\bar{N}_2\to \bar{d}X$. The amplitude
for such processes has been derived, e.g., from  models for the non-relativistic
nucleon-nucleon potential. As an alternative, Ref.~\cite{Dal:2015sha}
used experimental data to determine the cross sections
$\bar p \bar n\to \bar d X$ for $X=\{\gamma,\pi^0,\ldots\}$.
The coalescence probability was then determined as
$\sigma_{\rm tot}(\bar{N}_1\bar{N}_2\to \bar{d}X)/\sigma_0$ with $\sigma_0$
as a free
parameter. As a result, antideuterons were mainly produced with momenta
close to the delta resonance, $\sim 1$\,GeV, and the fit to the antideuteron
production data in $pp$ collision data from the ALICE experiment 
improved significantly.

In this work, we develop a coalescence model for the formation of light nuclei
in $e^+e^-$ and $pp$ collisions, which can be applied as well to dark matter
annihilations and decays.
Such a combined approach which describes successfully the production of light
nuclei both in ``point-like'' reactions ($e^+e^-$ , dark matter) and in $pp$ collisions is
especially needed for indirect dark matter searches,  where the consistent
prediction of ``signal'' antinuclei produced by dark matter annihilations and
of background antinuclei created in cosmic ray interactions is required.
Our model is based on the Wigner function
representations of the produced antinuclei states and allows us to include in
a semi-classical picture both the size of the formation region, which is
process dependent, and momentum correlations. The model contains a single,
universal parameter which is fixed by fitting the production
spectra of antideuterons in $pp$ collisions, measured by the ALICE experiment
at the LHC~\cite{Acharya:2017fvb}. The obtained value,
$\sigma_{(e^+e^-)}=\sigma_{(pp)}/\sqrt{2} \simeq 5\,{\rm GeV}^{-1} \simeq 1$\,fm,
agrees with its physical interpretation as the size of the formation
region of the light nuclei. Using this value, the model describes well the
data on the production of antihelium in $pp$ interactions and of antideuterons  
in $e^+e^-$ annihilations at the
$Z$-resonance~\cite{Schael:2006fd,Akers:1995az}.
%

\section{Wigner function based deuteron formation model}

We develop our model first for the case of deuteron production. The
generalization to helium-3 and tritium is straightforward and will
be performed in the next section. In the following, we use the fact
 that the binding energy $B$ of these nuclei is small, e.g.\ $B\simeq 2.2$\,MeV
for the deuteron. Therefore we can assume that a nucleus $A$ is formed
through the process $N_1+\ldots+N_n\to A^*$, and that the excitation energy
is later released by the emission of a photon.

\subsection{Derivation}

The starting point for the derivation of our new coalescence model is inspired
by the approach using Wigner functions, presented in Ref.~\cite{Scheibl:1998tk}.
We consider a system
consisting of a proton and a neutron in a frame where the motion of their CoM
is nonrelativistic. The number of deuterons with a given momentum $\vec P_d$
can be found by projecting the deuteron density matrix $\rho_d$ onto the
two-nucleon density matrix $\rho_\mathrm{nucl}$,
\begin{equation}
    \dv[3]{N_d}{P_d} = \tr{\rho_d\,\rho_\mathrm{nucl}}.
    \label{eq:proj_dens_mat}
\end{equation}
The deuteron density matrix describes a pure state,
$\rho_d = \ket{\phi_d}\bra{\phi_d}$. The spin and isospin values of the
two-nucleon state can be taken care of by introducing a statistical factor
$S=3/8$~\cite{Mattiello:1996gq}, such that the two-nucleon density matrix
can be written as $\rho_\mathrm{nucl}=\ket{\psi_p\psi_n}\bra{\psi_n\psi_p}$
and is normalized as
\begin{equation}
\bra{\psi_n\psi_p}\ket{\psi_p\psi_n}=N_p N_n  .
    \label{eq:rho-norm}
\end{equation}
Here, $N_p$ and  $N_n$ are the average multiplicities of protons and
neutrons per event, respectively.\footnote{We neglect for the moment the
double counting of nucleons involved in different pairs.}

By evaluating the trace in the coordinate representation
$\ket{\vec x_1\vec x_2}$, where the two
indices refer to the positions of the two nucleons, one finds
\begin{equation}
    \dv[3]{N_d}{P_d} = S\int\dd[3]{x_1}\dd[3]{x_2}\dd[3]{x_1'}\dd[3]{x_2'}\phi_d^*(\vec x_1, \vec x_2)\phi_d(\vec x_1', \vec x_2')\left<\psi^\dagger_n(\vec x_2')\psi^\dagger_p(\vec x_1')\psi_p(\vec x_1)\psi_n(\vec x_2)\right>,
    \label{eq:3.5_scheibl_heinz}
\end{equation}
where $\phi_d(\vec x_1, \vec x_2)$ and $\psi_i(\vec x)$ are the wave functions
of the deuteron and nucleon $i$, respectively. Next we factorise the deuteron
wave function into a plane wave describing the CoM motion with momentum
$\vec P_\mathrm{d}$ and an internal wave function $\varphi_d$,
\be
\phi_d(\vec x_1, \vec x_2) = (2\pi)^{-3/2}\exp{\iu\vec P_d \cdot(\vec x_1 + \vec x_2)/2} \varphi_d(\vec x_1 - \vec x_2) .
\ee
Then we replace the two-nucleon density matrix by its two-body Wigner function,
\begin{equation}
\begin{aligned}
  \left<\psi_n(\vec x_2')^\dagger\psi_p(\vec x_1')^\dagger\psi_p(\vec x_1)\psi_n(\vec x_2)\right>= &
  \int\frac{\dd[3]p_n}{(2\pi)^3}\frac{\dd[3]p_p}{(2\pi)^3}\,
  W_{np}\left(\vec p_n,\vec p_p, \frac{\vec x_2 + \vec x_2'}{2}, \frac{\vec x_1 + \vec x_1'}{2}\right)
  \\ & \quad\times \exp[\i \vec p_n\cdot (\vec x_2 - \vec x_2')]\exp[\i \vec p_p\cdot (\vec x_1 - \vec x_1')].
\label{eq:new_model_express_wigner}
\end{aligned}
\end{equation}
Further, we introduce as new coordinates the ``average'' positions of the proton
and neutron, $\vec r_p=(\vec x_1 +\vec x_1')/2$ and
$\vec r_n=(\vec x_2 +\vec x_2')/2$, as well as their separation
$\vec r=\vec r_n-\vec r_p$,
$\vec\xi=\vec x_1 -\vec x_1'-\vec x_2 +\vec x_2'$ and
$\vec\rho=(\vec x_1 -\vec x_1'+\vec x_2 -\vec x_2')/2$.
Changing also the momentum integration variables to  $\vec p= \vec p_n + \vec p_p$ and
$\vec q=(\vec p_n-\vec p_p)/2$, and performing then the $\vec\rho$ and $\vec p$ integrals,
we arrive at
\begin{equation}
    \dv[3]{N_d}{P_d} = \frac{S}{(2\pi)^6}\int\dd[3]{q} 
    \int \dd[3]{r_p}\dd[3]{r_n}\,\mathcal{D}(\vec r, \vec q)\,
    W_{np}(\vec P_d/2 +\vec{q}, \vec P_d/2 -\vec q, \vec{r}_n, \vec r_p),
    \label{eq:3.10new}
\end{equation}
where
\begin{equation}
  \mathcal{D}(\vec r, \vec q) =
  \int \dd[3]{\xi} \exp{-\i\vec q\cdot \vec \xi}
  \varphi_d(\vec r + \vec \xi/2)\varphi_d^*(\vec r - \vec \xi/2)
\label{Wd}
\end{equation}
is the Wigner function\footnote{Our conventions for the normalisation of the
  Wigner function are described in the Appendix~\protect\ref{appW}.}
of the internal deuteron wave function $\varphi_d$.

Using a Gaussian as ansatz for the deuteron wave function,
\begin{equation}
    \varphi_d(\vec r) = \left(\pi d^2\right)^{-3/4} \exp{-\frac{r^2}{2d^2}},
    \label{eq:ansatz_deuteron_wf}
\end{equation}
its Wigner function follows as
\begin{equation}
     \mathcal{D}(\vec r, \vec q) = 8 \me{-r^2/d^2}\me{-q^2d^2}.
     \label{eq:deutwignerinternal}
\end{equation}
The measured deuteron rms charge radius
$r_\mathrm{rms}=1.96$\,fm~\cite{Zhaba:2017syr} is reproduced
choosing\footnote{Note that the variable $r$ in the deuteron wave
  function describes the diameter, such that
  $r_\mathrm{rms}^2 = \int \dd[3]r (r/2)^2 |\varphi(r)|^2$.}
$d=3.2\unit{fm}$. To proceed, we have to choose also an ansatz for the
Wigner function of the two-nucleon state:
While Ref.~\cite{Scheibl:1998tk}  choses a thermal equilibrium state
motivated by the picture of a ``fireball'' formed in heavy-ion collisions,
we are in this work interested in the scattering of ``small'' systems as 
in $e^+e^-$, dark matter or $pp$ collisions. Therefore we use that
Monte Carlo simulations of strong interactions provide the momentum
distribution of the produced nucleons,
$G_{np}(\vec p_n,\vec p_p)$, which includes also  relevant momentum
correlations. On the other hand, $G_{np}(\vec p_n,\vec p_p)$  can be obtained
from the Wigner function as
\be
 \int{\rm d}^3r_p\,{\rm d}^3r_n \, W_{np}(\vec p_n,\vec p_p, \vec r_n,\vec r_p)
 =N_p N_n\, |\psi_{np}(\vec p_n,\vec p_p)|^2 \equiv G_{np}(\vec p_n,\vec p_p),
\ee
where $\psi_{np}(\vec p_n,\vec p_p)$ is the normalized two-nucleon wave
function in momentum space. We assume therefore a factorization of the
momentum and coordinate dependences,
\begin{equation}
W_{np}(\vec P_d/2 +\vec{q}, \vec P_d/2 -\vec q, \vec{r}_n, \vec r_p)  =
H_{np}(\vec{r}_n, \vec r_p) \, G_{np}(\vec P_d/2 +\vec{q}, \vec P_d/2 -\vec q).
\end{equation}
Note that this assumption implies a transition from a full quantum mechanical
treatment to a semi-classical picture.  Finally, we neglect spatial
correlations between the proton and the neutron,
$H_{np}(\vec{r}_n, \vec r_p)=h(\vec r_n)\,h(\vec r_p)$  and
choose a Gaussian ansatz for $h(\vec r)$,
\begin{equation}
    h(\vec r) =\left(2\pi\sigma^2\right)^{-3/2} \exp{-\frac{r^2}{2\sigma^2}}.
    \label{eq:ansatz_nucleon_wiger_function}
\end{equation}
Eq.~\eqref{eq:3.10new} then takes the form
\begin{equation}
    \dv[3]{N_d}{P_d} = \frac{3\zeta}{(2\pi)^6}  \int \dd[3]{q}\,\me{-q^2 d^2}\,
     G_{np}(\vec P_d/2 +\vec{q}, \vec P_d/2 -\vec q),
    \label{eq:new_model}
\end{equation}
where the factor
\begin{equation}
    \zeta \equiv \left(\frac{d^2}{d^2 + 4\sigma^2}\right)^{3/2}\leq 1
    \label{eq:new_model_zeta_1}
\end{equation}
depends on the characteristic spatial spread of the nucleons and on the spatial 
extension of the deuteron wave function. The coalescence probability is also
suppressed for large $q^2d^2$ as a Gaussian in our model.

\subsection{Parameter estimation}

In order to estimate the characteristic values for the parameter $\sigma$ in
the spatial distribution $h(\vec r)$ defined in
Eq.~(\ref{eq:ansatz_nucleon_wiger_function}), one generally has to consider 
separately the longitudinal and transverse directions, 
\begin{equation}
    h(\vec r)  \propto  \exp{-\frac{\vec r_\parallel^2}{2\sigma_\parallel^2}
     - \frac{\vec r_\perp^2}{2\sigma_\perp^2}}.
    \label{eq:new_wigner_nucelon_ansatz}
\end{equation}

\begin{figure*}[tb]
    \centering
    \includegraphics[width=0.8\textwidth]{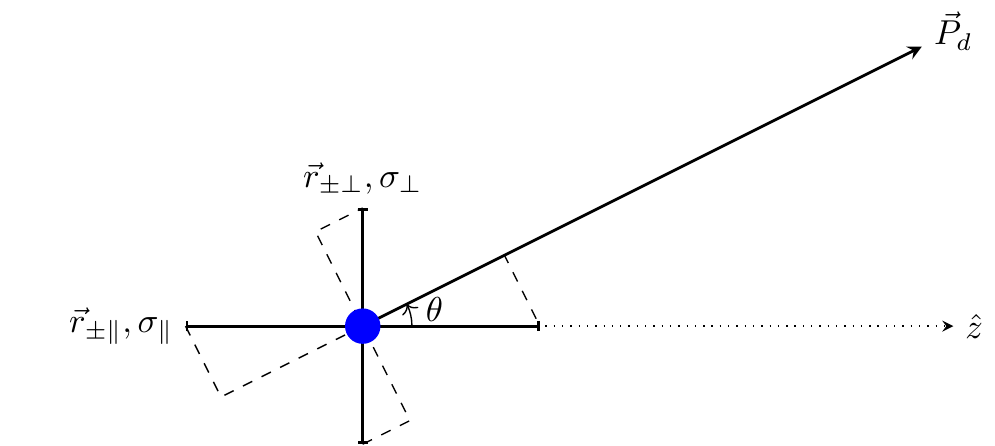}
    \caption{Splitting of the width $\sigma$ into a part parallel to the beam ($z$-axis), $\sigma_\parallel$, and a part perpendicular to the beam, $\sigma_\perp$. Only the components of $\vec r_\perp$ and $\vec r_\parallel$ parallel to the deuteron momentum $\vec P_d$ are affected by the Lorentz transformation. 
    \label{fig:lorentz_transform}}
\end{figure*}

Let us discuss first the case of $e^+e^-$ annihilation into hadrons in the
center-of-mass frame of the collision, choosing the $z$-axis along the 
direction of
the outgoing quark and antiquark, as shown in Fig.~\ref{fig:lorentz_transform}.
Before we proceed, it is worth remarking that this reaction involves
three different time and distance scales~\cite{Dokshitzer:1991wu}: 
The annihilation of the electron-positron
pair into the quark and antiquark happens during the time 
$t_{\rm ann}\sim 1/\sqrt{s}$.
For $s\gg  \Lambda_{\rm QCD}^2$,  the hard process is thus almost point-like
in coordinate space. The perturbative cascading of the produced (anti-) quark
proceeds via parton branchings with the characteristic momentum transfer squared
$\Lambda_{\rm QCD}^2\ll |q^2|\ll s$. This implies that the corresponding
longitudinal proper distance scales are smaller than $\Lambda_{\rm QCD}^{-1}$.
Therefore, the third and last step, the nonperturbative conversion of the final
partons into hadrons, corresponds to the longest time and distance scales:
The so-called  hadronisation time or formation length $L_{\rm had}$ required
for a hadron to build up its parton ``coat'' is
\begin{equation}
L_{\rm had} \sim \gamma L_0 ,
\label{l-form}
\end{equation}
where $\gamma$ is the gamma factor of the hadron in the considered frame and 
$L_0$ equals approximately the nucleon size, $L_0\sim R_{p} \sim 1$\,fm.
The coalescence process  involves nucleons which
have (almost) completed their formation and the process proceeds on distance
scales which are comparable to $L_{\rm had}$. Boosting to the rest frame
of the produced deuteron compensates the  gamma factor in Eq.~(\ref{l-form}),
hence we expect $\sigma_\parallel \sim L_0\sim 1$\,fm in that frame.

The characteristic transverse spread of a produced hadron can be estimated
using the uncertainty relation: The transverse displacement of the
hadron is obtained summing over the perpendicular components of the
random walk performed by the previous generations of partons during both
the perturbative and nonperturbative parton cascading. The contribution of
a single branching
is inversely proportional to the transverse momentum of the parton,
$\Delta b_i\sim 1/p_{\perp,i}$. Also here, nonperturbative physics gives
the dominating contribution with\footnote{The numerical value of
  $\Lambda_{\rm QCD}$ depends on the renormalisation scheme used and varies
  between 0.3 and 0.9\,GeV for three flavors~\cite{Deur:2016tte}.} $p_{\perp}\sim \Lambda_{\rm QCD}$.
Since the bulk of deuterons is produced with relatively small transverse
momenta, boosting to the rest frame has a small effect on 
$\sigma_{\perp}\sim \Lambda_{\rm QCD}^{-1}$. In the simplest option we
consider, we neglect therefore this boost. Since $\Lambda_{\rm QCD}^{-1}$ is
of the same order of magnitude as $L_0$, we set
in the following $\sigma_\parallel = \sigma_{\perp} = \sigma$,
to minimize the number of  parameters.
An alternative set-up which takes into account the effect of the transverse
boost on $\sigma_{\perp}$, will be discussed at the end of this subsection.

Let us now move to proton-proton, proton-nucleus and nucleus-nucleus
collisions. Here, the picture is modified by multiple scattering processes
involving multiparton interactions:  The proton and neutron taking part in
the coalescence process can thus originate from different parton-parton
interactions. Therefore it is necessary to take into account the longitudinal
and transverse spread of the initial parton clouds of the projectile and
the target. Starting
with the former, it is important to keep in mind that the effect of the
Lorentz contraction is different for fast and slow partons. At a given
rapidity $y$ in the laboratory system, partons from, say, the target
proton cloud are distributed over the longitudinal distance
$\sim R_p /\gamma =R_p\,e^{-y}$. Boosting to the deuteron rest-frame
compensates again the gamma factor, such that the resulting ``geometrical''
contribution  to $\sigma_\parallel$, due to the longitudinal extension 
of the parton cloud, equals $\sigma_{\parallel {\rm (geom)}}\sim R_p\sim 1$\,fm.
Summing the two contributions in quadrature, we obtain
\begin{equation}
 \sigma_{\parallel (pp)}^2 = \sigma_{\parallel(e^\pm)}^2 +
     \sigma_{\parallel(\mathrm{geom})}^2 \approx 2\sigma_{\parallel(e^\pm)}^2.
\end{equation}

\begin{figure}[tb]
    \centering
    \includegraphics[width=0.3\textwidth]{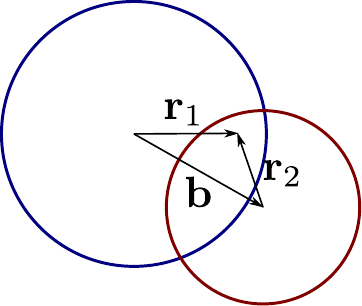}
    \caption{Sketch of the parton clouds of two interacting hadrons.}
    \label{fig:geom}
\end{figure}

Finally, we have to consider the geometrical contribution to $\sigma_\perp$.
One may naively expect it to depend on the impact parameter for a proton-proton
(proton-nucleus) collision. Let us show that this is not the case in the
simple geometrical picture of Fig.~\ref{fig:geom} and derive the geometrical
contribution to $\sigma_\perp$. We define $\sigma_{\perp(\mathrm{geom})}$ 
as the transverse spread of the overlapping region ($O$) of the 
projectile and target parton clouds,
\begin{equation}
\sigma_{(\mathrm{geom})}^2 = \langle r_1^2\rangle_O - \langle \vec r_1\rangle^2_O.
\end{equation}
The expectation value $\langle A\rangle_O$ follows then as
\begin{equation}
    \langle A\rangle = \frac{\int \dd[2]{r_1}\dd[2]{r_2}\,A\,
     \rho_1(r_1)\,\rho_2(r_2)\,
      w_\mathrm{int}(|\vec b-\vec r_1 + \vec r_2|)}
      {\int \dd[2]{r_1}\dd[2]{r_2}\, \rho_1(r_1)\,\rho_2(r_2)\,
       w_\mathrm{int}(|\vec b-\vec r_1 + \vec r_2|)},
\end{equation}
where $\vec b$ is the impact parameter for the collision, $\rho_i(r_i)$ are
the transverse parton densities of the projectile ($i=1$) and the target
($i=2$), and $w_\mathrm{int}$ is the probability for a parton-parton
interaction. Assuming for simplicity that the latter is point-like,
\begin{equation}
    w_\mathrm{int}(|\vec b-\vec r_1 + \vec r_2|)
    \propto \delta^{(2)}(\vec b-\vec r_1 + \vec r_2),
\end{equation}
and approximating the density distributions by Gaussians,
\begin{equation}
    \rho_i=1/(\pi\, R_i^2)\exp{-r^2/R_i^2},
    \label{eq:gaussian_parton_dist}
\end{equation}
with $R_i$ being the transverse radii of the projectile  and the target,
respectively, we obtain
\begin{equation}
 \sigma_{\perp(\mathrm{geom})}^2 = \frac{R_1^2R_2^2}{R_1^2+R_2^2}.
\label{eq:estimate_sigma}
\end{equation}
For the particular case of $pp$ collisions, we have
$\sigma_{\perp\mathrm{(geom)}}^2=R_p^2/2$.
Since this is of the same order of magnitude as $\sigma_{\perp(e^\pm)}^2$,
we set
\begin{equation}
    \sigma_{\perp (pp)}^2 = \sigma_{\perp(e^\pm)}^2 + \sigma_{\perp(\mathrm{geom})}^2 
    \approx 2\sigma_{\perp(e^\pm)}^2,
\label{eq:sig-perp-tot}
\end{equation}
such that we can also use for proton-proton collisions one universal
parameter,
\begin{equation}
 \sigma_{\parallel (pp)}  = \sigma_{\perp (pp)} = \sigma_{(pp)} =\sqrt{2}\,\sigma_{(e^\pm)}\,.
\label{eq:sigma-uni}
\end{equation}
It is noteworthy that such an assumption would  generally be
unjustified in the case of proton-nucleus and nucleus-nucleus
collisions since the corresponding geometrical contributions to
$\sigma_\parallel$ and $\sigma_{\perp}$ may differ significantly. 
Considering, as an example,
proton-lead collisions, we have\footnote{Note that the 
geometrical contributions to
$\sigma_\parallel$ and $\sigma_{\perp}$ should generally vary from event
to event, depending on the corresponding rate of multiple scatterings.
For proton-proton collisions, such a variation is expected to be relatively
weak and, hence may be neglected in a first approximation. This is,
however, different for  proton-nucleus and nucleus-nucleus interactions:
Our estimations are valid for relatively central collisions involving
numerous pair-wise inelastic rescatterings between the projectile and
target nucleons, which provide the bulk contribution to the formation of
(anti-)nuclei. On the other hand, 
 peripheral interactions at large impact parameters are dominated
by a single binary collision between a pair of projectile and target nucleons,
which gives rise to   $\sigma_{\parallel {\rm (geom)}}\sim R_p$,
like in the proton-proton case.  This leads to a pronounced correlation between 
the size of the source region
and the multiplicity of secondary hadrons produced.}
  $\sigma_{\parallel {\rm (geom)}}\sim R_{\rm Pb}$,
while Eq.~(\ref{eq:estimate_sigma}) yields for the  transverse 
spread
$\sigma_{\perp(\mathrm{geom})}\simeq R_p$.

\paragraph{Boosted \boldmath{$\sigma_\perp$}}

In an alternative set-up, we take into account that  $\sigma_{\perp}$ is
defined in the collider frame, while in the derivation of
Eq.~\eqref{eq:new_model}, all quantities and wave functions were evaluated
in the rest frame of the deuteron. This requires that we Lorentz transform
$W(\vec r,\vec q)\propto h(\vec r)$ between the two frames.
Such a transformation includes both a longitudinal boost, with a Lorentz
factor $\gamma_{\parallel}\simeq \gamma \cos \theta \simeq \gamma$, 
and a transverse one. Here
$\theta \simeq (p_{p\perp}+p_{n\perp})/(p_{p\parallel}+p_{n\parallel})$
is the small angle between the direction of motion of the nucleon pair
in the CoM and the $z$-axis in the original frame, before the boost.
While the former transformation 
has been accounted for in our definition for $\sigma_\parallel$, the effect
of the latter is to replace $\sigma_\perp$ defined in the original frame  
by $\tilde\sigma_\perp$ in the CoM, with
\be
 \tilde\sigma_\perp = \frac{\sigma_\perp}{\sqrt{\cos^2\theta + \gamma^2\sin^2\theta}} .
\label{eq:sig-tilde}
\ee
Thus, the factor $\zeta$ in Eq.\ (\ref{eq:new_model_zeta_1}) changes to
\begin{equation}
  \zeta = \frac{d^2}{d^2 +4 \tilde\sigma_\perp^2}
  \sqrt{\frac{d^2}{d^2 + 4\sigma_\parallel^2}}\,.
    \label{eq:new_model_final_2}
\end{equation}

\subsection{Numerical implementation}
\label{sec:num_proc}

As one can see in Eq.~(\ref{eq:new_model}), a given
proton-neutron pair with momentum  difference $2\vec q$ in its CoM has
the probability
\be
w =  3\zeta \e^{-q^2d^2}
\label{eq:w_ij}
\ee
to form a deuteron. Depending on whether we use the simplified approach or
take into account the modification of $\sigma_\perp$ by the transverse boost,
the factor $\zeta$ is defined by Eq.~(\ref{eq:new_model_zeta_1})
or~(\ref{eq:new_model_final_2}), respectively.

At this point, we have to take some care of potential double (triple, etc.)
counting since a given proton may be paired with different neutrons and vice
versa. Let us  assume that for a given event, the  final state contains
$N_p$ protons and $N_n$ neutrons. Denoting by $w_{ij}$ the coalescence 
probability, Eq.~(\ref{eq:w_ij}), for a pair formed out of the $i$-th proton
and the $j$-th neutron, we have the following expression for the average
number of deuterons produced in such an event,
\begin{align}
N_d =& \sum _{i=1}^{N_p} \sum _{j=1}^{N_n}w_{ij} 
       - \frac 12 \sum _{i=1}^{N_p} \sum _{k\neq i}^{N_p} \sum _{j=1}^{N_n}w_{ij}\,w_{kj}
-  \frac 12\sum _{i=1}^{N_p} \sum _{j=1}^{N_n} \sum _{l\neq j}^{N_n}w_{ij}\,w_{il}-\ldots  \nonumber \\
 \simeq &  \sum _{i=1}^{N_p} \sum _{j=1}^{N_n}w_{ij}
 \left[1- \frac 12\sum _{k\neq i}^{N_p}w_{kj} - \frac 12\sum _{l\neq j}^{N_n}w_{il}\right],
\label{eq:w_ij-corr}
\end{align}
where in the last line we have taken into account the 
smallness of the coalescence probabilities and have neglected the contributions
of  triple and higher contributions.

As one can see from  Eq.\ (\ref{eq:w_ij-corr}), the contribution of a given
proton-neutron pair~ $ij$ to the  binning of the deuteron spectrum should be
taken with the weight,
\begin{equation}
       \Omega_{ij}=w_{ij}\left[1- \frac 12 \sum _{k\neq i}^{N_p}w_{kj}
 - \frac 12 \sum _{l\neq j}^{N_n}w_{il}\right] .
\end{equation}
Since we bin the deuteron distribution ${\rm d}^3N_d /{\rm d}P_d^3$ in the
reference frame of the detector, no additional factor accounting for the
Lorentz transformation of the yield is necessary.

\subsection{Improving the deuteron wave function}

In the treatment above, a Gaussian which reproduces the measured $r_\mathrm{rms}$
value of the deuteron charge distribution was used as wave function.
However, it is known that the deuteron wave function is stronger peaked at
$r=0$ than a Gaussian. An alternative is the Hulthen wave function,
\begin{equation}
 \phi_d(\vec r) =\sqrt{\frac{ab(a+b)}{2\pi(a-b)^2}}\frac{\me{-ar}-\me{-br}}{r}
\label{eq:hulthen}
\end{equation}
with $a=0.23\,\mathrm{fm}^{-1}$ and $b=1.61\,\mathrm{fm}^{-1}$, which gives a
good description of the deuteron ground state~\cite{Zhaba:2017syr}.
Using this wave function, an analytical derivation of the weights would 
however be not possible. To obtain a better description of the deuteron, and
at the same time  to keep the problem analytical solvable, we use instead the
sum of two Gaussians as an ansatz for the deuteron wave function,
\begin{equation}
  \varphi_d(\vec r) = \pi^{-3/4}\left[\frac{\Delta^{1/2}}{d_1^{3/2}}\me{-r^2/(2d_1^2)} + \e^{\i\alpha} \frac{(1-\Delta)^{1/2}}{d_2^{3/2}}\me{-r^2/(2d_2^2)}\right],
\end{equation}
where we include a relative phase $\alpha$ between the two terms. 
Choosing $ \e^{\i\alpha}=\i$ leads to some simplifications.  First, the
probability distribution 
\begin{equation}
    |\varphi_d(\vec r)|^2 = \pi^{-3/2}\left[\frac{\Delta}{d_1^3}\me{-r^2/d_1^2} + \frac{1-\Delta}{d_2^3}\me{-r^2/d_2^2}\right] 
    \label{eq:deuttwpgaussprob}
\end{equation}
contains with this choice no mixed terms. Moreover, we will see below
that this choice leads to the same weight function as in the one-Gaussian
case.

\begin{figure}[htb]
    \centering
    \includegraphics[width=.7\textwidth]{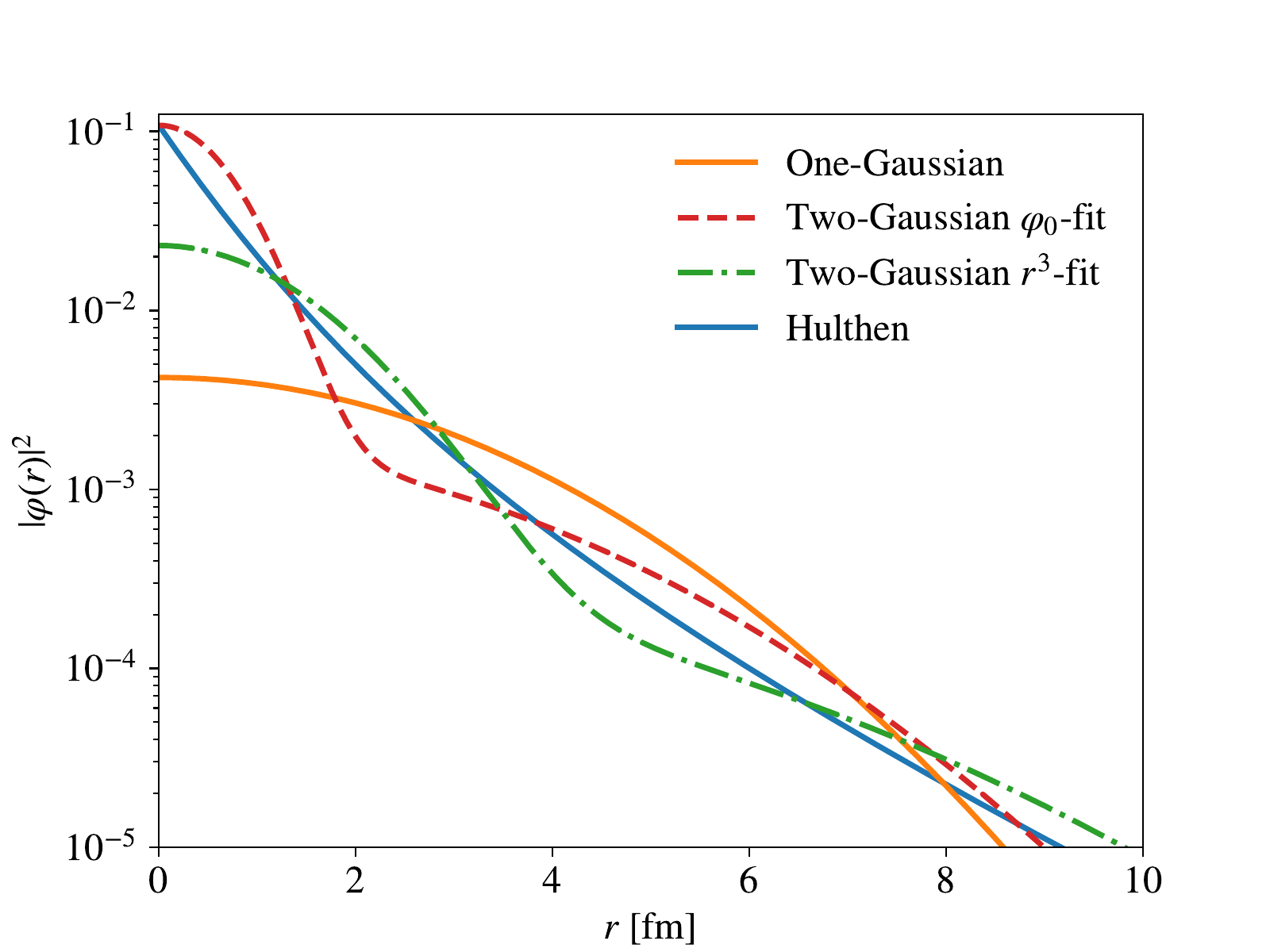}
    \caption{Comparison between the different parametrizations for the
    deuteron wave function.}
    \label{fig:new_model_wfs}
\end{figure}

Next we fit $|\varphi_d(\vec r)|^2$ to the Hulthen wave
function~\eqref{eq:hulthen} in order to fix
$\Delta$, $d_1$, and $d_2$. Two possible methods are to fit either
$|\varphi_d(0)|^2$, $\langle r\rangle$, and $\langle r^2\rangle$,  or 
$\langle r\rangle$, $\langle r^2\rangle$, and $\langle r^3\rangle$. The first
method will be called  $\varphi_0$-fit and the second $r^3$-fit. The
$\varphi_0$-fit yields $\Delta = 0.581$, $d_1 = 3.979$\,fm, and
$d_2=0.890$\,fm, while the $r^3$-fit yields $\Delta = 0.247$,
$d_1=5.343$\,fm, and $d_2=1.810$\,fm. The resulting probability distributions
are plotted in Fig.~\ref{fig:new_model_wfs} together with the one for the
one-Gaussian (Eq.~\eqref{eq:ansatz_deuteron_wf}) and the Hulthen
(Eq.~\eqref{eq:hulthen}) wave functions.  One can see in the figure that the
two-Gaussian ansatz resembles the Hulthen probability distribution more closely
than the  Gaussian wave function does, in particular, regarding the peak around
$r = 0$. The $\varphi_0$-fit reproduces visually the behaviour around $r=0$
best and will therefore be used when comparing with experimental data later on.

The deuteron Wigner function follows then as
\begin{equation}
    \mathcal{D}(\vec r, \vec q) = 8\left[ \Delta \me{-r^2/d_1^2}\me{-q^2d_1^2} + (1-\Delta)\me{-r^2/d_2^2}\me{-q^2d_2^2} \right] + A(\vec r \cdot\vec q),
\end{equation}
where the function $A$ is odd in $\vec r$,
$A(-\vec r \cdot\vec q)=-A(\vec r \cdot\vec q)$.
Therefore the new term  $A$ drops out performing the spatial integrals over
$\vec r_p$ and $\vec r_n$ in Eq.~(\ref{eq:3.10new}) and, thus,
does not contribute to the weights  in the binning procedure.
The  weights for the two-Gaussian case are thus
\begin{equation}
 w = 3\left(\zeta_1 \Delta e^{-q^2d_1^2} + \zeta_2[1-\Delta] e^{-q^2d_2^2}\right),
\end{equation}
where the $\zeta_i$ are given by Eq.~\eqref{eq:new_model_zeta_1}.

The weighted $q^2$ distributions with the one-Gaussian weight and
two-Gaussian weights for $pp$ collisions at $\sqrt{s}=7$\,TeV are
shown in Fig.~\ref{fig:q_distribution} for the ALICE setup discussed in
appendix~\ref{app:ALICE},  using  $\sigma=7$\,GeV$^{-1}$ and a constant
$\zeta$. For the old model, $p_0=0.2$\,GeV is used and the resulting
distribution is rescaled by a factor~0.3 to make the figure clearer.
Double counting has in all cases only a
minor effect on the resulting distributions. In the two-Gaussian case,
the better description of the peak at $r=0$ in the probability distribution
significantly enhances the contribution from  proton-neutron pairs with a
relatively large momentum  difference. 

\begin{figure}[htbp]
    \centering
    \includegraphics[width=0.9\textwidth]{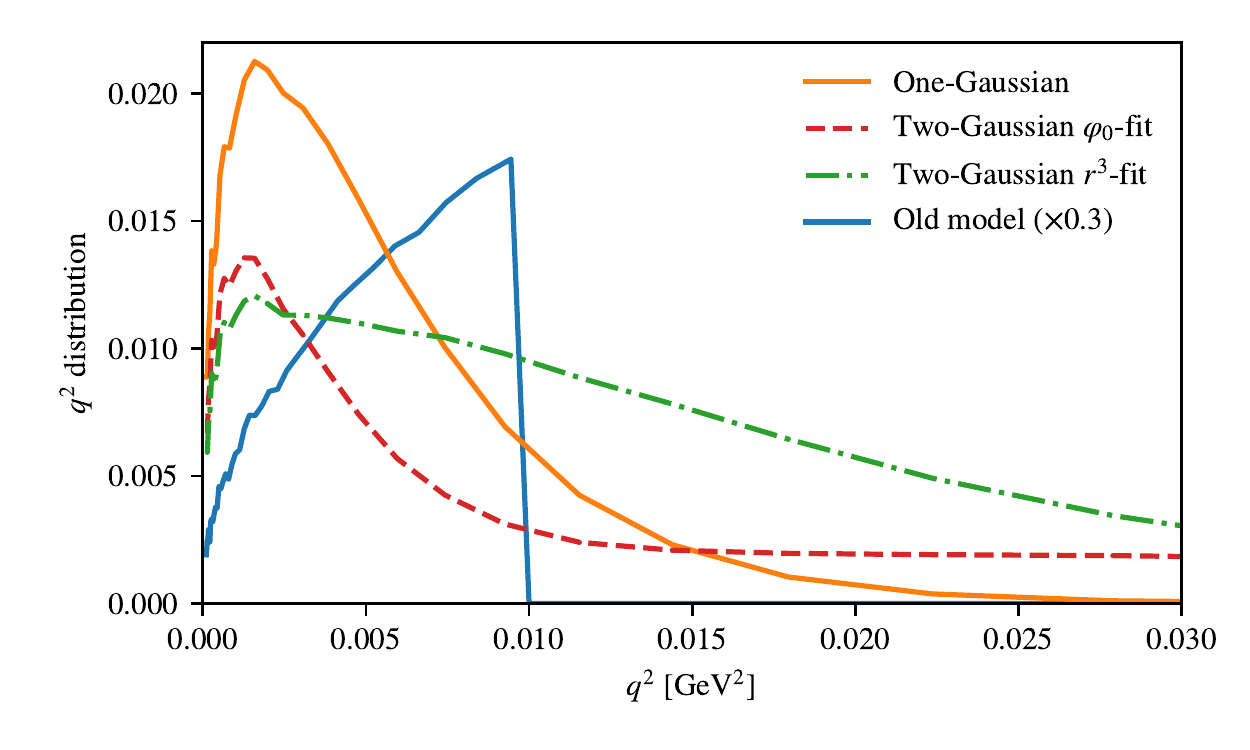}
    \caption{Weighted $q^2$-distribution for the four considered cases.
    \label{fig:q_distribution}}
\end{figure}

\section{Formation of helium-3 and tritium}

The cases of helium-3 and tritium nuclei are similar to the deuteron case,
but the derivation of the weight is more cumbersome. We account
  for the Coulomb interaction between the two protons in the helium nucleus
  only insofar as we allow for a different rms radius of the two nuclei in
  the fitting procedure. In this approach, our model applies in the same
  way for helium-3 and tritium.  Moreover, this assumption
  is supported by the data of the ALICE experiment which found a
  comparable yield of helium-3 and tritium nuclei. The binding energies of both
nuclei are still low ($\simeq 8$\,MeV), and the same approximations as in
the deuteron case thus still apply.

The number of helium nuclei with momentum $\vec P_\mathrm{He}$ is found by
projecting the helium density matrix onto the three-nucleon one, cf.\ with
Eq.~\eqref{eq:proj_dens_mat}. As in the deuteron case, the nucleus wave
function is factorised into a plane wave describing the CoM motion with
momentum $\vec P_\mathrm{He}$ and an internal wave function which depends
on the relative coordinates,
\begin{equation}
  \phi_\mathrm{He}(\vec x_1, \vec x_2, \vec x_3) = 
  (2\pi)^{-3/2} \exp{\iu\vec P_\mathrm{He}\cdot \vec R}\,
  \varphi_\mathrm{He}(\vec \rho,\vec  \lambda).
\end{equation}
Here, the Jacobi coordinates $\vec R$, $\vec\rho$, and $\vec\lambda$ are
expressed via $\vec x_1$,  $\vec x_2$ and $\vec x_3$ as
\begin{subequations}
    \begin{align}
        \vec \lambda &= (\vec x_1 + \vec x_2 - 2\vec x_3)/\sqrt 6, \\
        \vec \rho &= (\vec x_1 - \vec x_2)/\sqrt 2, \\
        \vec R &= (\vec x_1 +\vec x_2 + \vec x_3)/3,
    \end{align}
    \label{eq:new_model_helium_cm_coor}
\end{subequations}
with $\vec x_1^2 + \vec x_2^2 + \vec x_3^2 =
3\vec R^2 + \vec \rho^2 + \vec \lambda^2$, 
$\vec \rho^2 + \vec \lambda^2 = 
(\vec x_1-\vec x_2)^2+(\vec x_1-\vec x_3)^2+(\vec x_3-\vec x_2)^2$,
$\dd[3]{r_1}\dd[3]{r_2}\dd[3]{r_3} =
3^{3/2}\dd[3]{R}\dd[3]{\rho}\dd[3]{\lambda}$. 
The internal wave function is again approximated by a Gaussian in 
the relative coordinates $\vec\rho$ and $\vec\lambda$,
\begin{equation}
 \varphi_\mathrm{He}(\vec \rho,\vec  \lambda)  = (3\pi^2 b^4)^{-3/4}\,
    \exp{-\frac{\vec \rho^2 + \vec \lambda^2}{2b^2}},
\end{equation}
with $b$ being the rms radius  of the nucleus,
\begin{equation}
    r_\mathrm{rms}^2 = 3^{3/2}\int\dd[3]\rho\dd[3]{\lambda}\,
    \frac{\rho^2+\lambda^2}{3}\, |\varphi_\mathrm{He}(\vec \rho,\vec  \lambda)|^2=b^2.
\end{equation}
The $^3\mathrm{He}$ and $^3\mathrm{H}$ nuclei have  rms radii equal
1.96\,fm and 1.76\,fm, respectively~\cite{karshenboim_precision_2008}.

Performing the same steps as in Eqs.~(\ref{eq:3.5_scheibl_heinz})
and (\ref{eq:new_model_express_wigner}) in the deuteron case, we obtain for
the momentum spectrum of the produced nuclei
\begin{equation}
\begin{aligned}
  \dv[3]{N_\mathrm{He}}{P_\mathrm{He}} & =
\frac{S}{(2\pi)^3}\int \dd[3]{r_1}\dd[3]{r_2}\dd[3]{r_3}\dd[3]{r_1'}\dd[3]{r_2'}\dd[3]{r_3'}\,
 \me{-\iu\vec P_\mathrm{He}\cdot(\vec R - \vec R')}\,
  \varphi_\mathrm{He}(\vec \rho,\vec  \lambda)^*\,
  \varphi_\mathrm{He}(\vec \rho',\vec  \lambda')\\
    &\times \int\frac{\dd[3]{p_1}}{(2\pi)^3}\frac{\dd[3]{p_2}}{(2\pi)^3}
    \frac{\dd[3]{p_3}}{(2\pi)^3} \: 
    \me{\i\vec p_1\cdot(\vec x_1-\vec x_1')+\i\vec p_2\cdot(\vec x_2-\vec x_2')
    +\i\vec p_3\cdot(\vec x_3-\vec x_3')} \\
  &\times \: W_{N_1N_2N_3}\left(\vec p_1,\vec p_2,\vec p_3, \frac{\vec x_1 + \vec x_1'}{2}, 
  \frac{\vec x_2 + \vec x_2'}{2},   \frac{\vec x_3 + \vec x_3'}{2}\right),
 \end{aligned}
\end{equation}
where $S=1/12$ is the statistical factor accounting for the different isospin
and spin states and $W_{N_1N_2N_3}$ is the Wigner function for the three-nucleon
state. We approximate again $W_{N_1N_2N_3}$ by a product of momentum and 
coordinate distributions of the nucleons, neglecting spatial correlations
between the latter,
\begin{equation}
W_{N_1N_2N_3}(\vec p_1,\vec p_2,\vec p_3, \vec{r}_1, \vec r_2, \vec r_3)  =
  G_{N_1N_2N_3}(\vec p_1,\vec p_2,\vec p_3)\: \prod _{i=1}^3 h(\vec r_i),
\end{equation}
where  $h(\vec r)$ is given by Eq.~\eqref{eq:ansatz_nucleon_wiger_function}.

Expressing further the product 
$\varphi_\mathrm{He}(\vec \rho,\vec  \lambda)^*\,
  \varphi_\mathrm{He}(\vec \rho',\vec  \lambda')$ via the 
  Wigner function of the helium nucleus and doing the
spatial integrals, changing to the
coordinates~\eqref{eq:new_model_helium_cm_coor}, we finally obtain
\begin{equation}
    \dv[3]{N_\mathrm{He}}{P_\mathrm{He}}  = \frac{64S\, \zeta}{(2\pi)^9} \int
    \dd[3]{p_1}\dd[3]{p_2}\dd[3]{p_3}\, 
    \delta ^{(3)}(\vec p_1+\vec p_2 + \vec p_3-\vec P_\mathrm{He})\,
  G_{N_1N_2N_3}(\vec p_1,\vec p_2,\vec p_3)\,  \me{-b^2 P^2},
    \label{eq:new_model_final_1_helium}
\end{equation}
where
\begin{equation}
  \zeta = \left(\frac{b^2}{b^2 + 2\sigma^2}\right)^3
\end{equation}
accounts for the overlap of the wave functions and
\begin{equation}
    P^2=\frac{1}{3}\left[(\vec p_1 - \vec p_2)^2 + (\vec p_1-\vec p_3)^2 
    +(\vec p_2-\vec p_3)^2\right]
    \label{eq:new_model_helium_squared_momentum}
\end{equation}
is a measure of the relative momentum difference between the nucleons.
The procedure for finding the correct Lorentz transformation is similar to
the deuteron case. The result is
\begin{equation}
  \zeta = \left(\frac{b^2}{b^2 + 2\tilde\sigma_\perp^2}\right)^2
  \frac{b^2}{b^2 + 2\sigma_\parallel^2},
    \label{eq:new_model_zeta_helium}
\end{equation}
where $\tilde\sigma_\perp$ is again given by~\eqref{eq:new_model_final_2}.

The numerical procedure for treating the formation of  tritium and helium-3
nuclei is identical to the one described in section~\ref{sec:num_proc}, apart
from the different weight factor; $P^2$ is now defined in the CoM frame of
the three-particle state. 
One may argue that it is sufficient to calculate the momentum differences
between nucleons, entering Eq.~(\ref{eq:new_model_helium_squared_momentum}),
in the rest frames for the corresponding nucleon pairs since, because of
the exponential factor in Eq.~(\ref{eq:new_model_final_1_helium}),
those practically coincide with the ones defined in the 
rest frame of the nucleus.
This approach was used throughout this work.

\section{Comparison with experimental data}
\label{sec:comp_exp_data}

The predicted yield of antinuclei depends on the hadronisation
scheme~\cite{Dal:2012my}, and a comparison to different experimental data sets
should be therefore made using a single MC simulation. In this work,
we choose to perform all our simulations of $pp$ and  $e^+e^-$ collisions
with Pythia~8.230~\cite{Sjostrand:2006za,Sjostrand:2014zea} which describes
the antiproton spectra at LHC energies within 20--30\%~\cite{Adam:2015qaa}.
We set $\Delta \tau = 0$, switching thereby decays of long-lived particles
in Pythia off\footnote{This choice is similar to the approach of Ref.~\cite{Ibarra:2012cc}, as Pythia~8 only stores non-zero lifetimes in the event table that are relevant for displaced vertices in collider experiments.}.
Thereby we exclude nucleons which are produced mostly outside the
  source region.
In each run, we take into consideration all produced nucleon pairs 
with $q < 0.25$\,GeV for the  one-Gaussian case and with $q<0.5$\,GeV for the 
two-Gaussian ansatz for the deuteron wave function.
There are two obvious methods to generalise the standard per-event coalescence
model to helium-3 and tritium: One can require that each of the relative
momenta lie within a sphere with radius $p_0$ in momentum space, or that the
absolute momentum difference for each pair of particles is smaller than
$p_0$~\cite{Carlson:2014ssa,Cirelli:2014qia}. The latter approach was used here.

\begin{table*}
\centering
{\small
\begin{tabular}{@{}lllllll@{}}
\toprule
 Experiment & \multicolumn{2}{c}{one-Gaussian} & \multicolumn{2}{c}{two-Gaussian} & \multicolumn{2}{c}{Old model} \\
                & $\sigma$ [$\frac{1}{\mathrm{GeV}}$] &  $\frac{\chi^2}{N-1}$ & $\sigma$ [$\frac{1}{\mathrm{GeV}}$] &  $\frac{\chi^2}{N-1}$ & $p_0$ [MeV] &  $\frac{\chi^2}{N-1}$ \\
\midrule
ALICE 0.9\,TeV &  $3.5\pm 0.7$ & 7.5/2 & $6.2\pm 0.3$ & 6.0/2 & 181 & 7.3/2 \\
ALICE 2.76\,TeV & $4.3\pm 0.3$ & 44/6 & $6.6\pm 0.1$ & 32/6  & 174 & 45.6/6\\
ALICE 7\,TeV &  $4.1\pm 0.2$ & 182/19 & $6.6\pm 0.1$ & 133/19 & 176 & 177/19\\
ALICE combined & $4.1\pm 0.1$ & 235/29 & $6.6\pm 0.1$ & 172/29 & 176 & 229/19 \\
ALICE helium-3 & $4.5\pm 0.9$ & 1.7/2 & - & - & 179 & 1.2/2 \\
ALEPH & $0^{+2.3}_{-0}$ & - & $5.0^{+0.9}_{-0.6}$ & - & $214^{+21}_{-26}$ & -\\
ALEPH + OPAL & $0^{+4.4}_{-0}$ & 3.2/1 & $5.5^{+1.3}_{-1.1}$& 3.2/1 & 201 & 3.2/1\\
\bottomrule
\end{tabular}
}
\caption{Fit results for the constant $\zeta$ factor, in comparison to
 the old model.}
\label{tab:sigma_const}
\end{table*}

\begin{table}
\centering
{\small
\begin{tabular}{@{}lllll@{}}
\toprule
 Experiment & \multicolumn{2}{c}{one-Gaussian} & \multicolumn{2}{c}{two-Gaussian} \\
                & $\sigma$ [$\frac{1}{\mathrm{GeV}}$] &  $\frac{\chi^2}{N-1}$ & $\sigma$ [$\frac{1}{\mathrm{GeV}}$] &  $\frac{\chi^2}{N-1}$  \\
\midrule
ALICE 0.9\,TeV &  $3.9\pm 0.8$ & 6.7/2 & $6.9\pm 0.3$ & 2.6/2 \\
ALICE 2.76\,TeV & $4.9\pm 0.3$ & 35/6 & $7.5\pm 0.1$ & 8.6/6  \\
ALICE 7\,TeV &  $4.7\pm 0.2$ & 143/19 & $7.6\pm 0.1$ & 29/19  \\
ALICE combined & $4.7\pm 0.4$ & 186/29 & $7.6\pm 0.1$ & 45/29  \\
ALICE helium-3 & $5.2\pm 1.0$ & 1.1/2 & - & -  \\
ALEPH & $0^{+2.4}_{-0}$ & - & $5.3^{+1.0}_{-0.6}$ & - \\
ALEPH + OPAL & $0^{+4.6}_{-0}$ & 3.2/1 & $5.8^{+1.4}_{-1.1}$& 3.2/1 \\
\bottomrule
\end{tabular}
}
\caption{Fit results obtained taking into account the modification of
  $\sigma_\perp$ by transverse boosts.}
\label{tab:sigma_beam_dep}
\end{table}

Let us  now test our model with available data on antideuteron production in
$e^+e^-$ annihilation at the $Z$ resonance energy, from the ALEPH and OPAL
experiments, and  with ALICE data on antideuteron and antihelium production
in $pp$ collisions. Details of the experimental setups are described in
Appendix~B. Both  for the parameter $\sigma$ of the new model and 
for $p_0$ of the standard coalescence model, we perform   $\chi^2$  fits  to
these  data sets.
 The best-fit values, their 1\,$\sigma$ errors and the reduced
$\chi^2$ of the various fits are given in Table~\ref{tab:sigma_const} for
the case of constant $\sigma_\perp$. In turn, the fit  results  reported in 
Table~\ref{tab:sigma_beam_dep} take into account the modification of
$\sigma_\perp$ by transverse boosts, i.e., they have been obtained
using $\zeta$ defined in  Eq.~(\ref{eq:new_model_final_2}), with
$\tilde \sigma_\perp$ from Eq.~(\ref{eq:sig-tilde}).
We first note that the two-Gaussian cases lead to  significantly
reduced $\chi^2$ values, compared to the one-Gaussian ansatz or 
to the standard coalescence model. 
At the same time, they favor larger values for the
parameter $\sigma$,  which is related to an enhanced contribution from
nucleon pairs with relatively large $q^2$. Taking into account  
the modification of  $\sigma_\perp$ by transverse boosts improves 
significantly the quality of the fits, as one can see in 
Table~\ref{tab:sigma_beam_dep}.
Moreover, in that case, the  best-fit values of the parameter $\sigma$
agree well with our estimates in Section~2.2: The ratio of the 
values of  $\sigma$, determined from fits to  the ALICE data for  $pp$ 
collisions and to the ALEPH data for $e^+e^-$ annihilation,
agrees well with the expected one, equal $\sqrt{2}$ 
(c.f.\ Eq.\ (\ref{eq:sigma-uni})),
and the absolute value, $\sigma_{(e^+e^-)}\simeq 5\,{\rm GeV}^{-1}\simeq 1$\,fm,
is consistent with its interpretation as the characteristic hadronic length
scale ($\sim R_p$).

\begin{figure*}[tb]
    \centering
    \includegraphics[width=\textwidth]{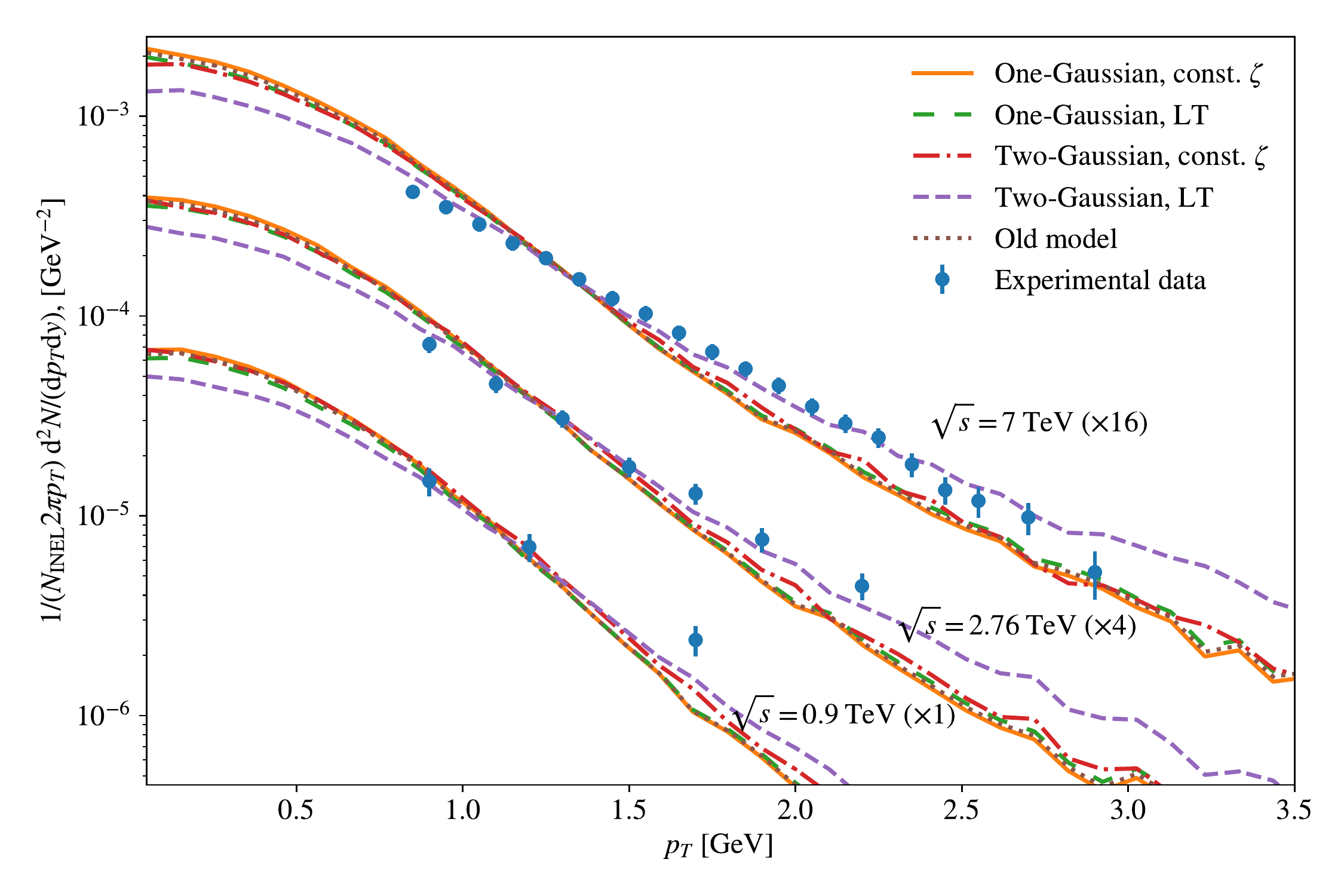}
    \caption{Best combined fits to the ALICE antideuteron data for the considered models. The data and fits are multiplied by a constant factor to make the figure clearer. The curves labelled LT are obtained iincluding the Lorentz boost of $\sigma_\perp$.
    \label{fig:alice_plot}}
\end{figure*}

\begin{figure*}[tb]
    \centering
    \includegraphics[width=0.95\textwidth]{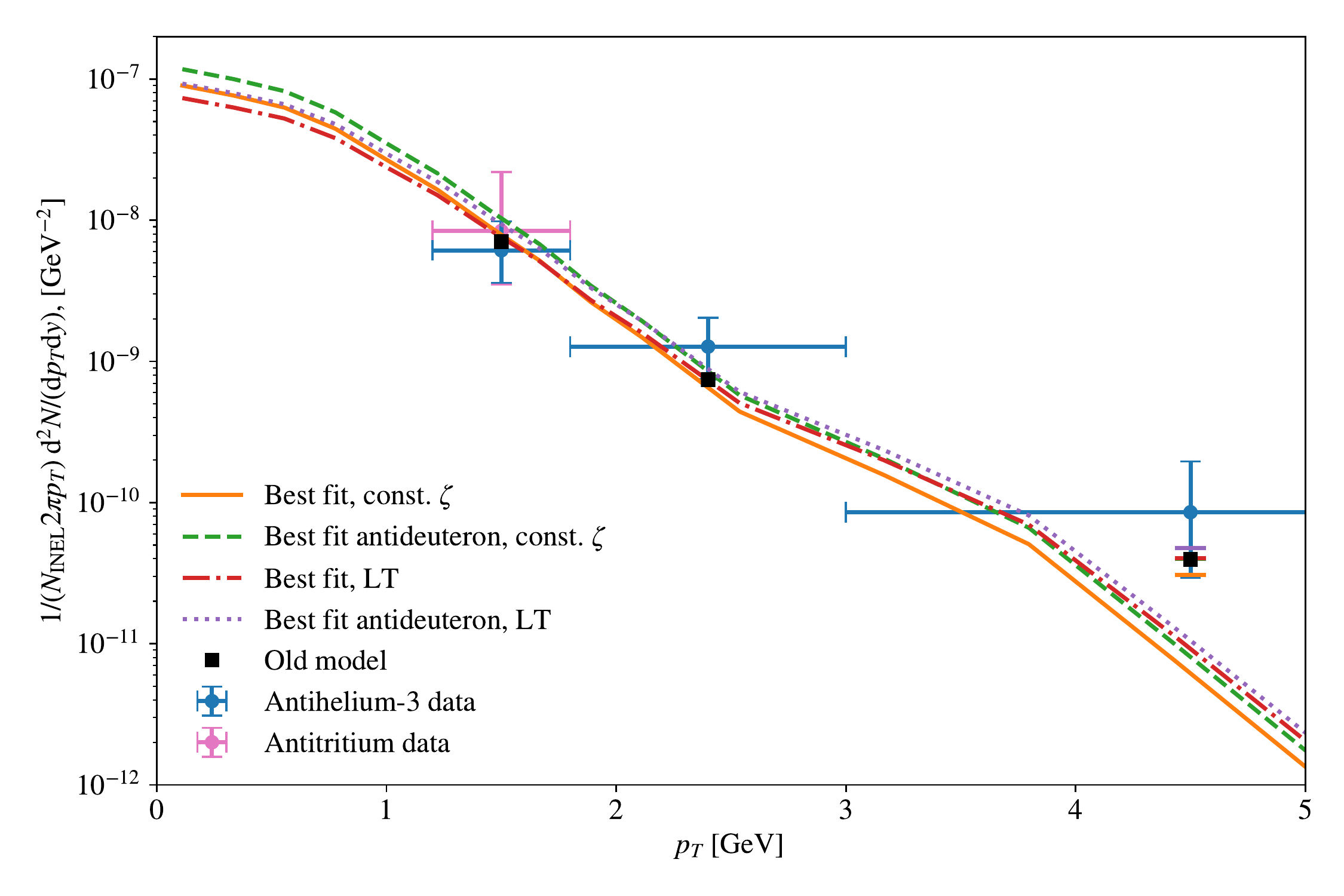}
    \caption{Best fits to the ALICE antihelium data for the one-Gaussian models. The best fits using the parameters obtained from the best combined fit to the ALICE antideuteron data is also plotted.}
    \label{fig:alice_helium_plot}
\end{figure*}

In Fig.~\ref{fig:alice_plot}, we compare the ALICE antideuteron data with
the best-fit spectra obtained for the various cases considered. It
becomes apparent that the slope of the $p_\perp$ distributions is best
described using the two-Gaussian ansatz for the deuteron wave function,
when the modification of $\sigma_\perp$ by transverse boosts is taken into
account.

In Fig.~\ref{fig:alice_helium_plot}, we compare the
predictions of our model to the ALICE data for light antinuclei,
antihelium and a single data point for tritium. Using the best-fit values
for $\sigma$ obtained from fitting the  antideuteron data in the
two-Gaussian case, the experimental data are satisfactorily reproduced.
Additionally, we show  the best-fit spectra obtained by fitting the
antihelium data. Since the difference between the two fits compared to
the errorbars is small, we conclude that the two data sets can be consistently
described using the same value of $\sigma$.
The goodness-of-fit parameter is $\chi^2/(N-1)=6.0/2$, when the $\sigma$
determined in the combined antideuteron fit is used for the helium-3 and
tritium data.
Thus there is a slight tension, and it will be interesting to  check it by
future antihelium data with reduced errors.

\section{Conclusion}

We have developed a new coalescence model for the formation of antinuclei, 
which combines an event-by-event Monte Carlo description of the collision
process with a microscopic coalescence   treatment based on the Wigner
function representations of the produced antinuclei states. This approach
has allowed us to include in a semi-classical picture both the size
$\sigma$ of the formation region and the momentum correlations of the
nucleons forming the nuclei. Since the size $\sigma$ is process dependent,
the difference in the observed antideuterons yields in $e^+e^-$ and $pp$
collisions can be naturally accounted for. Fitting the single, universal
parameter contained in our model to experimental data, we obtained
a best-fit value, $\sigma\simeq 1$\,fm, which corresponds well to its
physical interpretation as the size of the formation region of the light
nuclei.
If in the future,  antideuteron and  antihelium data sets with reduced
errors and for a larger $p_\perp$ range will be available, an independent
fit of the two parameters $\sigma_\perp$ and $\sigma_\|$ might be warranted.

We have examined different approximations for the deuteron wave functions
as well as two different implementations of the transverse size
$\sigma_\perp$ of the formation region. The fits to the ALICE antideuteron data prefer the two-Gaussian
wave function and the approach where  the effect on $\sigma_\perp$, due to a
 Lorentz boost to the deuteron frame, is taken into account. 
 Both  correspond to the physically expected
choices: The two-Gaussian wave function takes into account that the
deuteron wave function is rather peaked at $r=0$, while
$\sigma_\perp\simeq R_p$ is expected to hold in the CoM frame
of the collider. Using the best-fit values for antideuteron, we could
describe well the data for the production of antihelium in $pp$ interactions
and of antideuterons in $e^+e^-$ annihilation at the $Z$-resonance energy.

Our model is therefore well suited to investigate in detail the antideuteron
and antihelium fluxes predicted in models for dark matter annihilations and
from cosmic ray interactions. In particular, it will be interesting to see
whether and how the tentative antihelium events announced by the AMS-02
collaboration~\cite{ting16} can be explained within our model.

\section*{Acknowledgements}
\noindent
M.K.\ and J.T.\ acknowledge partial support from the Research Council of
Norway (NFR).
S.O.\ acknowledges support from project OS\,481/2-1 of the 
Deutsche Forschungsgemeinschaft.

\appendix

\section{Wigner function}                               \label{appW}

Our definition~(\ref{Wd}) of the one-particle Wigner function implies in
$d=1$ as normalisation (with $\hbar = 1= h/(2\pi)$)
\be 
\int \frac{{\rm d}p}{2\pi} {\rm d}x \,W(x,p)= 1 \,.
\ee
The corresponding probability distributions for the space and momentum
variables are obtained from
\begin{eqnarray} 
\int {\rm d}x \,W(x,p)=\psi^*(p)\,\psi(p), \\
\int\frac{{\rm d}p}{2\pi} \,W(x,p)=\phi^*(x)\,\phi(x) .
\end{eqnarray}  
For our ansatz $W(x,p) = h(x)g(p)$, it follows that $h(x)$ describes the
probability distribution of the nucleon in coordinate space, while  the
probability distribution of the nucleon momenta $g(p)$ is normalised as
\be
\int \frac{{\rm d}p}{2\pi} g(p)= 1 \,.
\label{wp}
\ee

\section{Experiments}

\subsection{ALICE}
\label{app:ALICE}

The ALICE Collaboration measured the invariant differential yields of
deuterons and antideuterons,
\begin{equation}
 E\,\dv[3]{n}{p}=\frac{1}{N_\mathrm{inel}}\frac{1}{2\pi p_T}\dvv{N}{p_T}{y},
\end{equation}
in inelastic proton-proton collisions at centre of
mass energies $\sqrt{s} = 0.9, 2.76$ and 7\,TeV, in the $p_T$ range
$0.8<p_T<3$\,GeV and for rapidity\footnote{Note that while an additional
  pseudo-rapidity cut $|\eta|<0.8$ was required in the data selection,
  the measurements have been  corrected to correspond to the  $|y|<0.5$
  selection, including also the contribution of $|\eta|>0.8$,
  using a MC simulation~\cite{Acharya:2017fvb}.} 
$|y|<0.5$~\cite{Acharya:2017fvb}. Here $E$ and $\vec p$ are the deuteron
energy and  momentum, $N_\mathrm{inel}$ is the number of inelastic events,
$N$ is the total number of detected deuterons, and $n\equiv N/N_\mathrm{inel}$. 
The experiment included a trigger (V0) consisting of two hodoscopes of 32
scintillators that covered the pseudo-rapidity ranges $2.8<\eta<5.1$ and
$-3.7<\eta < -1.7$, used to select Non-Diffractive (ND) inelastic events.
An event was triggered by requiring a hit (charged particle) on either 
side (positive or negative $\eta$) of the V0 triggering setup.

Pythia 8 generates general inelastic collisions, including single-diffractive
(SD), double-diffractive (DD) and ND events. The minimum bias events selected
by the V0 trigger generally include those that Pythia  treats as
SD and DD events. While we used Pythia 8 to generate general  minimum 
bias $pp$ collisions,  only events satisfying the V0 trigger have been
included in our analysis.

\subsection{ALEPH and OPAL}
\label{app:ALEPH_OPAL}

The ALEPH collaboration at LEP studied the deuteron and antideuteron 
production in $e^+e^-$ collisions at the $Z$ resonance energy. The
measured production rate of antideuterons was
$(5.9\pm 1.8 \pm 0.5)\times 10^{-6}$ per
hadronic $Z$ decay, for the antideuteron momentum range from 
0.62 to 1.03\,GeV and for the production angle $\theta$ satisfying
$|\cos\theta|<0.95$~\cite{Schael:2006fd}.

In a similar experiment performed by the OPAL collaboration~\cite{Akers:1995az},
no antideuteron events were detected. Reference~\cite{Dal:2015sha}
noted that the resulting upper limit on the antideuteron yield has previously
been neglected, but should also be taken into account. The measurements were
performed in the antideuteron momentum range $0.35 < p < 1.1$\,GeV, with an
estimated detection efficiency $\epsilon=0.234$, which includes the angular
acceptance. The expected total number of antideuterons was
\begin{equation}
    N_{\bar d} = \epsilon N_\mathrm{ev} n_{\bar d, \mathrm{MC}},
\end{equation}
where $N_\mathrm{ev}=1.64\times 10^6$ is the number of events  in the
OPAL analysis and $n_{\bar d, \mathrm{MC}}$ is the MC prediction for the number
of antideuterons per event. We follow Ref.~\cite{Dal:2015sha} and
assume a Poissonian uncertainty $\sigma_{\bar d}=\sqrt{N_{\bar d}}$ 
for the expected number of antideuterons.
The $\chi^2$ is in this case given by
\begin{equation}
    \chi_\mathrm{OPAL}^2 = \frac{(N_\mathrm{obs} - N_{\bar d})^2}{\sigma_{\bar d}^2}
     = N_{\bar d} .
\end{equation}


\end{document}